\documentclass[a4paper,11pt]{article}
\usepackage{jcappub} 
\usepackage{natbib}
\usepackage{comment}
\usepackage{float}
\usepackage{multirow}
\usepackage{siunitx}
\usepackage{verbatim}
\usepackage{graphicx}
\usepackage{amsmath}
\usepackage[usenames]{color}
\usepackage{amssymb}
\usepackage{hyperref}
\usepackage{cleveref}
\newcommand{\be}{\begin{equation}}
\newcommand{\ee}{\end{equation}}
\newcommand{\bea}{\begin{aligned}}
\newcommand{\eea}{\end{aligned}}
\newcommand{\pr}{\partial}

\newcommand{\bse}{\begin{subequations}}
\newcommand{\ese}{\end{subequations}}

\newcommand{\bmm}{\begin{multline}}
\newcommand{\emm}{\end{multline}}

\begin{document}
\title{Analytical solution of traversable wormholes in the presence of positive cosmological constant}
\author[a]{Rajesh Karmakar}
\emailAdd{rajesh@shu.edu.cn}
\affiliation[a,b]{Department of Physics, College of Sciences, Shanghai University,
99 Shangda Road, 200444 Shanghai, China}
\author[b]{Xian-Hui Ge}
\emailAdd{gexh@shu.edu.cn}

\abstract 
{The construction of traversable wormholes with a cosmological constant, $\Lambda$, introduces significant challenges and leads to non-trivial modifications of the spacetime geometry. In this work, we obtain an analytical solution describing a locally traversable wormhole for $\Lambda>0$ following a perturbative approach. Starting from the Ellis–Bronnikov wormhole geometry, we derive the metric deformation induced by the cosmological constant at linear order, assuming $\Lambda r^2_0\ll 1$, where $r_0$ represents the throat of the Ellis–Bronnikov wormhole. We further identify the radial domain of validity of the perturbative solution, which is restricted to $r\ll\sqrt{3/\Lambda}$. Within this domain, we find that the Ellis–Bronnikov wormhole is modified by the cosmological constant; interestingly, the shape function takes a de Sitter-like form. 
Nevertheless, we verify that the flare-out condition holds at the deformed throat and find a violation of the null energy condition in its vicinity, as required for traversable WHs. Traversability is further analyzed by evaluating tidal forces and deriving constraints on the velocity required for safe human passage. Additionally, we discuss the possibility of constructing a global extension by matching the perturbative wormhole solution to a Schwarzschild-de Sitter exterior spacetime. Given the recent advances in the observational study of astrophysical compact objects, together with a variety of astrophysical and cosmological observations indicating the existence of a positive cosmological constant, the analytical solution presented here may be of considerable phenomenological interest.}

\maketitle

\section{Introduction}\label{intro}
The concept of a wormhole (WH) geometry has a long history, dating back nearly a century \cite{flamm1916beitrage, Einstein:1935tc}. It was later put on a systematic footing by Ellis, Morris, and Thorne through the development of the idea of traversability into a well-defined physical framework \cite{Ellis1973EtherFT, morris1988wormholes}. Since then, WH spacetimes have been extensively investigated in a variety of contexts (see, e.g., \cite{Bueno:2017hyj, Cardoso:2016rao, Churilova:2019qph, churilova2021wormholes} and references therein). A key motivation for studying such geometries is the replacement of spacetime singularities with a regular, traversable throat. Hence, the usual notion of the cosmic censorship conjecture is extended for WH spacetime and known as WH cosmic censorship \cite{del2019wormhole, bixano2025brief}. Concerning the global structure, the two asymptotic regions connected by the throat may correspond either to separate universes or distant regions of the same universe, offering a natural setting for questions related to entanglement and information transfer \cite{gao2017traversable, Cao:2024xaw}. Although no observational evidence for such geometries has been found so far, their potential signatures and astrophysical implications have been explored in recent years \cite{DeSimone:2025sgu, Azad:2022qqn, DeFalco:2021btn, DeFalco:2020afv, Blazquez-Salcedo:2018ipc, dai2019observing, Vagnozzi:2022moj, Riley:2026avn}.

A central challenge in constructing traversable WH solutions is the requirement of matter sources that violate the null energy condition (NEC), in contrast to standard gravitating matter, to sustain the throat against collapse \cite{morris1988wormholes, churilova2021wormholes, Maldacena:2018lmt}. Essentially, the energy condition-violating matter is required to produce the defocusing of null geodesic congruences, which in turn ensures the flare-out condition at the WH throat \cite{hawking2023large}. Interestingly, in some extended gravity theories, where the geodesic convergence condition is easily violated, people have resorted to constructing the WH solution without the need for such matter \cite{Blazquez-Salcedo:2020czn, konoplya2022traversable, Kanti:2011jz, antoniou2020novel}. Nevertheless, some form of matter is almost always present to support the WH, which makes the Einstein equations especially difficult to solve, often leaving numerical methods as the only viable approach. However, for purposes of applicability and phenomenological investigation, it is generally preferable to have analytical solutions. For certain specific matter sources, it has indeed been possible to obtain analytical forms of traversable WH solutions. For example, it has been found long ago that a phantom-like scalar field with negative kinetic energy can source the traversable WHs \cite{Armendariz-Picon:2002gjc}. Later on, traversable WHs have been constructed in Einstein GR and various modified gravity theories (see, e.g., \cite{loewer2024study, canate2022novel, maldacena2021humanly, hochberg1997geometric, simmonds2025traversable, Alshal:2019owh, DeFalco:2021ksd, Chew:2021vxh, Maldacena:2018gjk, Fernando:2016ksb, Richarte:2009zz, kanti2012stable, hochberg1997geometric}, and references therein). In this context, it is important to investigate how a non-vanishing cosmological constant, particularly a positive one, modifies or supports WH geometries.

A variety of cosmological observations, such as measurements of anisotropies in the cosmic microwave background \cite{Planck:2018vyg}, the tension in the Hubble parameter \cite{Riess:2019cxk}, and gravitational lensing of active galactic nuclei \cite{DES:2021wwk, Denzel:2020zuq}, indicate the presence of a dark energy component of the universe with a repulsive nature, commonly described by a positive cosmological constant, $\Lambda > 0$. If such a uniform background source is present, it will modify the asymptotic nature of the geometry of the gravitating objects. Previously, by introducing the cosmological constant through the cosmological scale-factor in the Morris–Thorne type geometry, an evolving WH solution was derived, supported by a phantom scalar field \cite{cataldo2009evolving} (for a rigorous discussion on the WH solution in the expanding cosmological background see \cite{KordZangeneh:2025qel}). In contrast, in the present work, we focus on the static configurations. 

Over the years, several attempts have been made to construct static WH solutions in the presence of a cosmological constant. Regarding analytical constructions, previous approaches have mainly considered matching an asymptotically flat WH interior to a Schwarzschild–de-Sitter exterior spacetime \cite{lemos2003morris}. In such constructions, the effect of the cosmological constant enters primarily through the exterior geometry and the matching procedure at the junction surface, while its role in the wormhole interior is reflected mainly in reducing the degree of {\it exoticity} required in the supporting matter source. In another construction, an analytical de Sitter-WH solution was obtained by introducing a matter shell between two black holes in a de Sitter background \cite{Dai:2018vrw}. A more straightforward treatment of WH supported by a scalar field in de Sitter spacetime was presented in \cite{Anabalon:2012tu}. Recently, charged WH configurations in de Sitter spacetime have been investigated using an ansatz inspired again by the Schwarzschild–de Sitter metric with an additional charge term \cite{Kim:2025zyo}. In the present work, however, we do not adopt such an ansatz. Instead, we use an existing analytical solution describing an asymptotically flat WH geometry and study its deformation in the presence of the cosmological constant. 

Given a homogeneous cosmological constant background, it is reasonable to assume that the spacetime retains its spherical symmetry. Then, we follow a perturbative approach to study the deformation induced by the presence of the cosmological constant in a seed solution, which, for the present case, is taken to be an asymptotically flat, spherically symmetric, static WH. For the sake of simplifying the analysis, we have considered the deformation up to a linear order. The perturbative approach to solving Einstein equations, particularly when considering deviations from vacuum solutions, has been widely implemented in various contexts, such as in deriving higher-curvature corrections to vacuum black hole solutions \cite{Cano:2019ore, Kanti:1995vq}. Nevertheless, we implement this perturbative approach in the presence of a cosmological constant to derive the corresponding deformation of an asymptotically flat seed WH geometry.

The remainder of the paper is organized as follows. In \Cref {sec: gd_review}, we briefly outline the procedure of the perturbative framework in a generic manner. Then, we apply this method to an asymptotically flat WH solution in \Cref{sec: ds_soln} to obtain the deformed solutions. The throat position of such WHs, flare-out conditions, the violation of the null-energy condition, and the traversability through the WH throat have also been analyzed in this section. Afterwards, the radial geodesics for massless particles have been investigated. In \Cref{sec: exterior}, we discuss the possibility of globally extending the locally deformed EB WH solution by matching it to an exterior spacetime. Finally, in \Cref {sec: conclusion}, we conclude with a future outlook of the present work.
\section{The perturbative decoupling of the source term}\label{sec: gd_review}
In this section, we develop a perturbative framework to decouple an additional source term in the Einstein equations\footnote {For a non-perturbative approach to decoupling source terms, see, e.g., \cite{Ovalle:2017fgl, Ovalle:2018vmg}.}. This approach should apply to generic static, spherically symmetric WH spacetimes. We begin by considering a seed solution, described by a static spherically symmetric spacetime as follows,
\be\label{eq:gen_seed_soln}
ds^2=-e^{\xi(r)}dt^2+e^{\mu(r)}dr^2+r^2d\theta^2+r^2\sin^2\theta d\varphi^2,
\ee
where $\xi(r)$ and $\mu(r)$ are redshift and shape function, respectively, and considered to be arbitrary functions of the radial coordinate for the moment. If such a spacetime is supported by a non-zero source term ($T_{\mu\nu}$), as is usual in the case of a WH, it should satisfy the Einstein equation, expressed as, 
\be
G_{\mu\nu}\equiv R_{\mu\nu}-\frac{1}{2}Rg_{\mu\nu}=\kappa^2T_{\mu\nu},
\ee
where $\kappa^2\equiv 8\pi G$, stands for the surface gravity term. Substituting the line element \eqref{eq:gen_seed_soln}, the non-zero components of the Einstein equation turn out as,
\be\label{eq:seed_EE}
\bea
&{G_0}^0=-\frac{1}{r^2}+e^{-\mu}\left(\frac{1}{r^2}-\frac{\mu'}{r}\right),\\
&{G_1}^1=-\frac{1}{r^2}+e^{-\mu}\left(\frac{1}{r^2}+\frac{\xi'}{r}\right),\\
&{G_2}^2=\frac{e^{-\mu}}{4}\left(2\xi''+\xi'^2-\mu'\xi'+2\frac{\xi'-\mu'}{r}\right).
\eea
\ee
Here and throughout the analysis, the prime denotes the derivative with respect to $r$, unless otherwise specified. These equations encompass the known seed solution, which serves as the starting point for the analysis. In the presence of an additional source the Einstein equation modifies to
\be\label{eq:modified_EE}
\tilde{G}_{\mu\nu}\equiv \tilde{R}_{\mu\nu}-\frac{1}{2}\tilde{R}g_{\mu\nu}=\kappa^2T_{\mu\nu}+\alpha S_{\mu\nu}
\ee
where $\alpha$ serves as a counting parameter, as will become clear in the discussion below, and is ultimately eliminated at the end of the analysis. In the presence of the extra source term, $S_{\mu\nu}$, keeping the spherical symmetry intact, we consider the following ansatz form of the solution of \eqref{eq:modified_EE},
\be\label{eq.2ndseed}
ds^2=-e^{\nu(r)}dt^2+e^{\sigma(r)}dr^2+r^2d\theta^2+r^2\sin^2\theta d\varphi^2,
\ee
where $\nu(r)$ and $\sigma(r)$ are arbitrary functions of the radial coordinate, and to be determined through the GD analysis. Substituting this line element, the non-zero components of the Einstein equation \eqref{eq:modified_EE} reads 
\be\label{eq: 2nd_modified_EE}
\bea
&{\Tilde{G}_0}{}^0=-\frac{1}{r^2}+e^{-\sigma}\left(\frac{1}{r^2}-\frac{\sigma'}{r}\right),\\
&{\Tilde{G}_1}{}^1=-\frac{1}{r^2}+e^{-\sigma}\left(\frac{1}{r^2}+\frac{\nu'}{r}\right),\\
&{\Tilde{G}_2}{}^2=\frac{e^{-\sigma}}{2}\left(\nu''+\frac{\nu'^2}{2}-\sigma'\nu'+\frac{\nu'-\sigma'}{r}\right).
\eea
\ee
In what follows, we assume that the background geometry is minimally modified and implement a perturbative expansion of the metric functions in terms of the cosmological constant. Accordingly, the deformations of the redshift and shape functions are considered only up to linear order, $\mathcal{O}(\alpha)$ and are given by\footnote{In the literature on gravitational decoupling (GD) \cite{Ovalle:2023ref, hua2025nonsingular, Contreras:2021yxe, Ovalle:2020kpd, tello2023gravitationally}, particularly in the context of black hole solutions, a similar form for deformation is usually considered; however, without treating the deformation as a perturbation.} 
\be
\bea
&\nu(r)\simeq \xi(r)+\alpha \eta_1(r)+\mathcal{O}(\alpha^2)\\
&e^{-\sigma(r)}\simeq e^{-\mu(r)}+\alpha \eta_2(r)+\mathcal{O}(\alpha^2),
\eea
\ee
where $\eta_1(r)$ and $\eta_2(r)$ are the function of $r$, to be determined. Therefore, we are now left with the determination of these unknown functions. Substituting these deformed functions in \eqref{eq: 2nd_modified_EE}, and linearizing the Einstein equations about $\alpha$, and keeping only the terms up to $\mathcal{O}(\alpha)$, we get:
\be\label{eq:seedplusgd}
\bea
-\frac{1}{r^2}+e^{-\mu}\left(\frac{1}{r^2}-\frac{\mu'}{r}\right)+\alpha  \left(\frac{\eta_2}{r^2}+\frac{\eta'_2}{r}\right)&\simeq\kappa^2{T_0}^0+\alpha {S_0}^0,\\
-\frac{1}{r^2}+e^{-\mu}\left(\frac{1}{r^2}+\frac{\xi'}{r}\right)+\alpha\left[\eta_2 \left(\frac{1}{r^2}+\frac{\xi '(r)}{r}\right)+\frac{e^{-\mu}\eta'_1}{r}\right]&\simeq\kappa^2{T_1}^1+\alpha {S_1}^1,\\
\frac{e^{-\mu}}{4}\left(2\xi''+\xi'^2-\mu'\xi'+2\frac{\xi'-\mu'}{r}\right)+\alpha \frac{e^{-\mu}\left(2r\eta''_1-r\eta'_1\mu'+2r\eta'_1\xi'+2\eta'_1\right)}{4 r}&\\
+\frac{\alpha}{4 r}\left(r\eta'_2\xi'+2\eta'_2+2r\eta_2\xi''+r\eta_2\xi'^2+2\eta_2\xi'\right)&\simeq\kappa^2{T_2}^2+\alpha {S_2}^2.
\eea
\ee
It is important to note that the above perturbative setup is implemented without imposing an explicit restriction on the cosmological constant at this stage. Once the deformation functions are obtained, the self-consistency of the approximation will be checked by verifying that the $\Lambda$-dependent corrections remain subdominant compared with the seed geometry within the physically relevant region. Following the above set of equations, it turns out that one can effectively write, up to $\mathcal{O}(\alpha)$,
\be
\tilde{G}_{\mu}{}^\nu\simeq {G_\mu}^\nu+\alpha{\mathcal{G}_\mu}^\nu,
\ee 
where $\mathcal{G}_{\mu\nu}$, arising due to the effective source term involving $S_{\mu\nu}$, as can be realized from \eqref{eq:seedplusgd}, and using \eqref{eq:seed_EE}. Nevertheless, the decoupled equation governing the deformation reads
\be\label{eq: decoupled_EE}
\bea
\left(\frac{\eta_2}{r^2}+\frac{\eta'_2}{r}\right)&\simeq {S_0}^0,\\
\left[\eta_2 \left(\frac{1}{r^2}+\frac{\xi '(r)}{r}\right)+\frac{e^{-\mu}\eta'_1}{r}\right]&\simeq {S_1}^1,\\
\frac{e^{-\mu}\left(2r\eta''_1-r\eta'_1\mu'+2r\eta'_1\xi'+2\eta'_1\right)}{4 r}+\frac{e^{-\mu}}{4}\left(2\xi''+\xi'^2-\mu'\xi'+2\frac{\xi'-\mu'}{r}\right)&\\
+\frac{1}{4 r}\left(r\eta'_2\xi'+2\eta'_2+2r\eta_2\xi''+r\eta_2\xi'^2+2\eta_2\xi'\right)
&\simeq {S_2}^2.
\eea
\ee
The above set of equations provides a general framework for computing the deformation of the seed solution \eqref{eq:gen_seed_soln}. In the next section, we will consider an existing analytical form of an asymptotically flat traversable WH spacetime and use it as the seed solution for our analysis. 
\section{Application to existing spherically symmetric asymptotically flat wormholes} \label{sec: ds_soln}
We consider the WB WH, which belongs to the Morris-Thorne category, as the seed solution to study the deformation. This WH is supported by a phantom-like scalar field \cite{Ellis1973EtherFT, Bronnikov:1973fh}, described by a minimally coupled massless scalar field. With such a matter field, the Einstein-Hilbert action is given by
\be\label{eq: EB_action}
\mathcal{A}=\int \sqrt{-g}d^4x\left[\frac{1}{16\pi G}R-\left(-\frac{1}{2}\pr_\mu\phi\pr^\mu \phi\right)\right].
\ee
The Einstein equation corresponding to this action turns out as
\be
G_{\mu\nu}\equiv R_{\mu\nu}-\frac{1}{2}Rg_{\mu\nu}=\kappa^2T^{(\phi)}_{\mu\nu},
\ee
where $T^{(\phi)}_{\mu\nu}$ stands for the stress-energy tensor of the massless scalar field, expressed as
\be
T^{(\phi)}_{\mu\nu}=-\pr_\mu\phi\pr_\nu\phi+\frac{1}{2}g_{\mu\nu}\pr_\alpha\phi\pr^\alpha\phi.
\ee
Notably, the unusual sign in the above expression is due to the negative kinetic term of the scalar field describing the phantom field. Nevertheless, the solution generated from this setup is asymptotically flat and spherically symmetric, and known as  Ellis-Bronnikov WH \cite{Ellis1973EtherFT, Bronnikov:1973fh}, described by the line element
\be\label{eq:MT_seed_soln}
ds^2=-dt^2+\frac{dr^2}{1-\frac{b(r)}{r}}+r^2d\theta^2+r^2\sin^2\theta d\varphi^2,
\ee
where, in comparison to \eqref{eq:gen_seed_soln}, we have 
\be
\xi(r)=0,~~\mu(r)=-\ln\left[1-\frac{b(r)}{r}\right],
\ee
with $b(r)=r^2_0/r$. Where $r_0$ denotes the WH throat, and the flare out condition $b'(r)|_{r=r_0}<1$ is guaranteed from the given expression.
To study the deformation in the spacetime for the positive cosmological constant, $\Lambda$, the above solution will be treated as the seed solution for the GD analysis. In the presence of $\Lambda$ the action \eqref{eq: EB_action} reads
\be\label{eq: EB_Lambda_action}
\mathcal{A}^{(\Lambda)}=\int \sqrt{-g}d^4x\left[\frac{1}{16\pi G}\left(R-2\Lambda\right)+\frac{1}{2}\pr_\mu\phi\pr^\mu \phi\right].
\ee
The corresponding Einstein equation can be expressed as
\be\label{eq:modified_lambda_EE}
\tilde{G}_{\mu\nu}\equiv \tilde{R}_{\mu\nu}-\frac{1}{2}\tilde{R}g_{\mu\nu}=\kappa^2T^{(\phi)}_{\mu\nu}-\alpha \Lambda g_{\mu\nu},
\ee
where $\alpha$ serves as a counting parameter as mentioned previously, and is ultimately eliminated at the end of the analysis. From the discussion of the previous section, the equations governing the unknown functions characterizing the deformations, up to $\mathcal{O}(\alpha)$, read (see \eqref{eq: decoupled_EE})   
\be\label{eq: deform_eqns}
\bea
\frac{\eta_2}{r^2}+\frac{\eta'_2}{r}&\simeq-\Lambda,\\
\frac{\eta_2}{r^2}+\frac{e^{-\mu}\eta'_1}{r}&\simeq-\Lambda,\\
e^{-\mu}\frac{\left(2r\eta''_1-r\eta'_1\mu'+2\eta'_1\right)}{4 r}+\frac{\eta'_2}{2 r}
&\simeq -\Lambda.
\eea
\ee
It is important to note that, even for the case $\xi(r)=0$, the perturbative expansion should be understood at the level of the metric coefficients (which enter into the Einstein equation) rather than the redshift function itself. In particular, the relevant expansion is that of $e^{\xi(r)+\alpha\eta_1(r)}$ , which reduces to $e^{\alpha\eta_1(r)}$ for the present seed geometry. Nevertheless, considering the first equation from the above set we obtain
\be
\bea
&\frac{d(r\eta_2)}{dr}\simeq-r^2\Lambda
\implies & \eta_2\simeq-\frac{r^2\Lambda}{3}+\frac{C}{r},
\eea
\ee
where $C$ is an arbitrary constant to be fixed. Whereas the second equation of \eqref{eq: deform_eqns} provides
\be
-\frac{\Lambda}{3}+\frac{C}{r^3}+\left(1-\frac{r^2_0}{r^2}\right)\frac{\eta'_1(r)}{r}\simeq-\Lambda.
\ee
The general solution of this equation turns out to be
\be
\eta_1(r)\simeq -\frac{\Lambda}{3}\,r^2-\frac{\Lambda r_0^2}{3}\,\ln\!\left|\frac{r^2}{r_0^2}-1\right|
-\frac{C}{2 r_0}\,\ln\!\left|\frac{r-r_0}{r+r_0}\right|+D,
\ee
where $D$ is another unknown constant.
To fix the unknown constants we proceed by inserting the first and second equations from the set \eqref{eq: deform_eqns} in the third one by replacing $\eta_2$, so that the resulting equation reads
\be
2r\eta''_1-r\eta'_1\mu'+4\eta'_1
\simeq 4re^{\mu}\Lambda.
\ee
Substituting the expression of $\eta_1$ from the previous equation in the above, we find that the above equation holds for $C=0$, with arbitrary $D$. For simplification, we fix the arbitrary constant $D=0$. Having determined the deformation functions, $\eta_1$ and $\eta_2$, the Ellis-Bronnikov WH spacetime in the presence of a cosmological constant can be expressed as 
\be\label{eq:dsEBmetric}
\bea
ds^2 &=-e^{-\frac{\Lambda}{3}\,r^2-\frac{\Lambda r_0^2}{3}\,\ln\!\left|\frac{r^2}{r_0^2}-1\right|}dt^2+\frac{dr^2}{1-\frac{r^2_0}{r^2}-\frac{r^2\Lambda}{3}}+r^2(d\theta^2+\sin^2\theta d\varphi^2).
\eea
\ee
Notably, the redshift function contains a Logarithmic divergence at $r=r_0$, therefore, the line element suitably describes the WHs, which have throat $r_{\rm th}> r_0$. To determine the throat of the WH, let us look at the root of the shape function, which can be expressed as \footnote{From the expression of the roots, it is straightforward to realize that for $\Lambda<0$, only the root $r_-$ is physical. In comparison to the asymptotically flat case, where the throat is located at $r_0$, we find that the throat radius shrinks for AdS. This aspect is analogous to a shrinking horizon size in the case of an asymptotically AdS black hole \cite{stuchlik1999some}. However, given that $r_-<r_0$, the redshift function in the line element can become logarithmically divergent as it crosses $r=r_0$. Whereas for a suitable WH geometry, the redshift function should be everywhere finite \cite{morris1988wormholes}. For this reason, we avoid discussing the AdS case in the present article, and we will investigate these issues further in our future work. For existing solutions with a negative cosmological constant and the motivations behind them, see, e.g., \cite{Blazquez-Salcedo:2020nsa}.},
\be
r_{\pm}=\sqrt{\frac{3}{2}\left(\frac{1\pm\sqrt{1-\frac{4}{3}\Lambda r^2_0}}{\Lambda}\right)}.
\ee
\begin{figure}[t!]
    \centering
\includegraphics[scale=0.55]{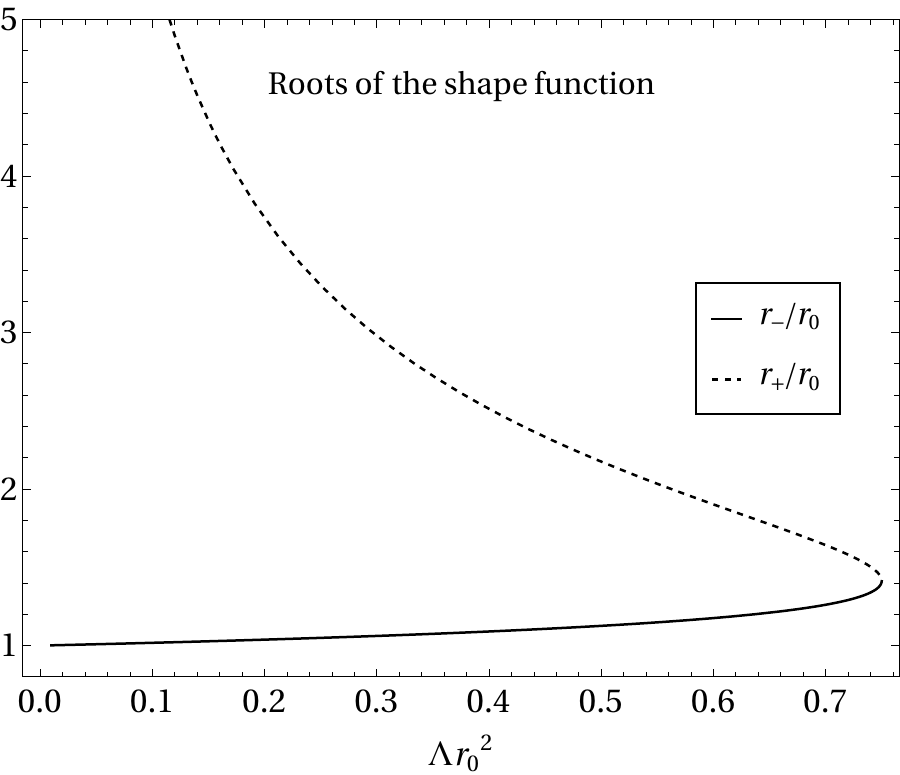}
\caption{The variation of the two roots of the shape function, $g^{rr}$, for the deformed EB WH geometry in the presence of the cosmological constant. The plus and minus signs are the inner and outer roots, respectively.}\label{fig: ds_wh_throat}
\end{figure}
For de Sitter spacetime with $\Lambda>0$, the physically motivated choice indicates that both roots are admissible, provided that $\Lambda\leq \Lambda_{\rm max}$, where $\Lambda_{\rm max}=3/(4r^2_0)$. Importantly, a closer inspection shows that the negative root, $r^{-}$, approaches the radius $r_0$ in the limit $\Lambda\to 0$ from above. Now, the positive root $r_+$, scaling as $1/\sqrt{\Lambda}$, is pushed to a large radius when $\Lambda\to 0$.  On the other hand, as $\Lambda\to \Lambda_{\rm max}$, the two roots, $r_{\pm}$, converge to $\sqrt{2}r_0$ as illustrated in the Fig.~\ref{fig: ds_wh_throat}. From this figure, it is also evident that the root positions always remain outside $r_0$. Having obtained the root locations for $\Lambda \leq \Lambda_{\rm max}$, it is now essential to verify the consistency of the perturbative expansion. For this purpose, let us first check the deformation of the redshift function. In this regard, the condition for the perturbative deformation to be subdominant with respect to the seed geometry, and thereby, the condition for the validity of the linear order approximation can be translated to
\be
\left|-\frac{\Lambda}{3}\,r^2-\frac{\Lambda r_0^2}{3}\,\ln\!\left|\frac{r^2}{r_0^2}-1\right|\right|\ll 1.
\ee
This relation imposes the conditions $r\ll\sqrt{3/\Lambda}$ and $\Lambda r^2_0\ll 1$. Importantly, as long as the above condition is satisfied, $g_{tt}$ remains finite and perturbative. It immediately suggests that the outer root, $r_+$, which scales as $1/\sqrt{\Lambda}$, does not satisfy the above criterion in the required sense. Likewise, a similar condition for the shape function can be quantified as 
\be
\frac{r^2\Lambda}{3}\ll\left|1-\frac{r^2_0}{r^2}\right|.
\ee
It should be emphasized, however, that the above criterion is not expected to hold at the root location itself. Since the roots are determined by the zeros of the deformed radial metric function, the zeroth-order and first-order contributions necessarily become comparable, as is generally encountered in perturbative constructions of gravitational solutions, such as perturbative corrections to black hole geometries \cite{Cano:2019ore, Kanti:1995vq}. Away from the roots, however, the perturbative expansion remains valid as long as $\Lambda r_0^2\ll1$ and $r\ll\sqrt{3/\Lambda}$. Importantly, for phenomenological applications in a de Sitter background \cite{Ovalle:2021jzf}, the values of the cosmological constant considered in the literature (see, e.g., \cite{stuchlik1999some, Ovalle:2021jzf}) satisfy $\Lambda r_0^2\ll1$ for astrophysically relevant throat scales. Consequently, the perturbative treatment adopted here remains well justified. The above discussion suggests that the inner root $r_-$ should be the potential location for the throat, where the redshift also remains finite. Let us, therefore, investigate further the properties of the inner root to be a traversable throat\footnote{See \cite{Kim:2025zyo} for a discussion of identifying the outer root as the cosmological throat. In our framework, since we find the outer root to be lying beyond the validity of the linear-order perturbative approximation, it is therefore left for future investigation whether a full non-perturbative analysis could further address this issue.}.
\begin{figure}[t!]
    \centering
\includegraphics[scale=0.7]{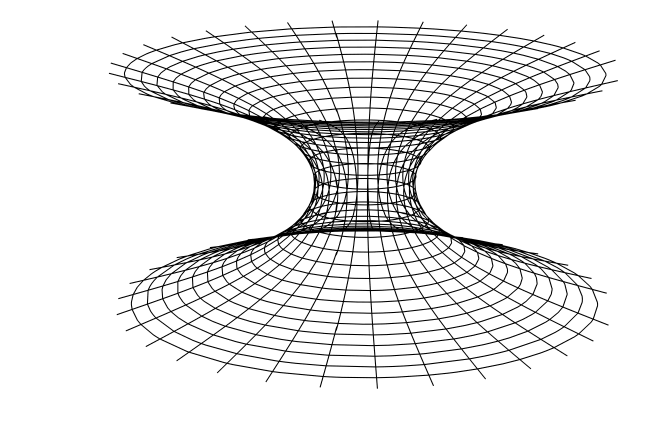}
\caption{The embedding diagram of the two dimensional section ($t$-constant, $\theta=\pi/2$) of the deformed EB WH in 3D Euclidean space (in cylindrical coordinate). Note that the plot is generated with $\Lambda=0.05$ and $r_0=1$. For visual clarity, we have not shown the effect due to the variation of the $\Lambda$ here, which is already understood from the previous figure \ref{fig: ds_wh_throat}. It is evident that the near-throat geometry of the two-dimensional spatial surface at the equatorial plane is very much similar to the standard asymptotically flat Morris-Thorne type WHs.}\label{fig: wh_fig}
\end{figure}

\begin{figure}[t!]
    \centering
\includegraphics[scale=0.55]{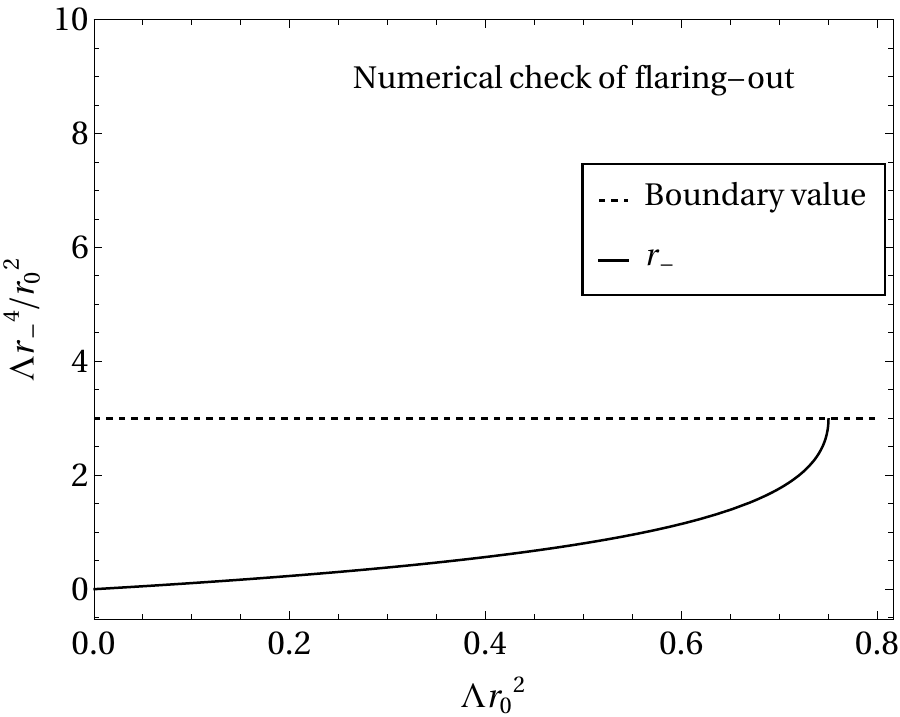}
\caption{The quantity $\Lambda r_-^4/r_0^2$ is shown for $\Lambda r_0^2<1$. The dotted horizontal line represents the critical value obtained from the flare-out condition. For the deformed EB wormhole throat $r_-$, the quantity $\Lambda r_-^4/r_0^2$ remains below this threshold, thereby satisfying the flare out condition discussed in the main text.}\label{fig: flare_out}
\end{figure}

A key requirement for a traversable WH geometry is the so-called flare-out condition, which can be derived by embedding the WH in a constant time slice. In comparison with the Morris–Thorne-type geometry \cite{morris1988wormholes}, we obtain the following relation for the present deformed EB WH:
\be
1-\frac{b(r)}{r}=1-\frac{r^2_0}{r^2}-\frac{r^2\Lambda}{3}.
\ee
Having spherical symmetry, it is suitable to restrict ourselves to the equatorial plane, $\theta=\pi/2$, so that the line element \eqref{eq:dsEBmetric}, in terms of the above formulation, takes the following form:
\be
ds^2=\left(1-\frac{b(r)}{r}\right)^{-1}dr^2+r^2d\phi^2.
\ee
Therefore, the geometry can be easily embedded into Euclidean space, $ds^2=dz^2+dr^2+r^2d\phi^2$, with which we obtain,
\be
ds^2=\left[1+\left(\frac{dz}{dr}\right)^2\right]dr^2+r^2d\phi^2,
\ee
with
\be
\frac{dz}{dr}=\pm\frac{1}{\sqrt{\frac{r}{b(r)}-1}}.
\ee
In Fig.~\ref{fig: wh_fig}, we have presented the embedding diagram of the above two-dimensional spatial surface of the deformed EB WH \eqref{eq:dsEBmetric} in the equatorial plane. Nevertheless, the flare-out condition is obtained by the requirement of the minimality of the WH throat \cite{Kim:2013tsa}, near $r_-$, which can be quantified as 
\be\label{eq: flare_out_cond}
\frac{d}{dz}\left(\frac{dr}{dz}\right)=\frac{b(r)- r b'(r)}{b(r)^2}
=\frac{\frac{2r_0^2}{r} - \frac{2\Lambda r^3}{3}}
{\left(\frac{r_0^2}{r} + \frac{\Lambda r^3}{3}\right)^2}>0
\ee
This condition, further, translates to
\be
\frac{\Lambda r^4}{r^2_0}<3.
\ee
In Fig.~\ref{fig: flare_out}, we have illustrated how this condition is satisfied for the throat located at $r_-$.

Another equivalent approach to realize the existence of the throat and thereby the flaring out is to utilize the proper radial distance defined as 
\be
dl=\frac{dr}{\sqrt{1-\frac{r^2_0}{r^2}-\frac{r^2\Lambda}{3}}}.
\ee
The deformed EB WH line element in terms of the above proper radial distance reads
\be
ds^2=-e^{-\frac{\Lambda}{3}\,r^2(l)-\frac{\Lambda r_0^2}{3}\,\ln\!\left|\frac{r^2(l)}{r_0^2}-1\right|}dt^2+dl^2+r^2(l)(d\theta^2+\sin^2d\varphi^2).
\ee
Then the throat of the WH is characterized by the extrema in the areal radius (or, equivalently, of the area of the two-sphere) of the above line element. Quantitatively, it can be expressed as
\be
\frac{dr}{dl}=\sqrt{1-\frac{r^2_0}{r^2}-\frac{r^2\Lambda}{3}}=0.
\ee
This obviously coincides with the root of $g^{-1}_{rr}$ and therefore the existence of the local minima at $r_-$. Furthermore, the flare out condition requires that the extrema correspond to a local minimum of the areal radius, namely, $d^2r/dl^2>0$. Using the definition of the proper radial distance, we obtain
\be
\frac{d^2r}{dl^2}=\frac{r_0^2}{r^3(l)} - \frac{\Lambda r(l)}{3}.
\ee
This condition is equivalent to \eqref{eq: flare_out_cond} and establishes $r_-$ as the viable throat with flaring-out property. Nevertheless, the proper radial distance formulation makes explicit that the local minimum of the areal radius of the spherical sector of the spacetime intrinsically characterizes the traversable WH throat.  In the subsequent discussion of this section, we will investigate whether the present deformed EB WH solution is stable and assess whether it is viable for traversability.
\subsection{Violation of the null energy conditions}
So far, following the GD approach, the cosmological constant term is treated as an effective source term here. However, it is conceivable that this strategy is essentially implemented for computational simplicity, allowing for the analytical solution to be obtained. The physical source term for the asymptotically flat WHs, which we considered for the seed solution, remains the actual source supporting the final deformed EB WH. Hence, the violation of the null energy condition (which implies the violation of the weak and strong energy conditions as well) for the asymptotically flat case is already checked. Now, with the null vectors of the deformed EB WH spacetime, let us check the null energy condition, which is defined as
\be
T_{\mu\nu}k^\mu k^\nu \geq 0
\ee
where $T_{\mu\nu}$ represents the stress-energy tensor for the asymptotically flat WHs, which, as per our discussion, still supports the de Sitter extensions. Whereas $k^\nu$ represents the null vector, which satisfies $k_\mu k^\mu=0$. One possible choice for a null vector, having spherical symmetry in the spacetime, is a radial null vector
\be
k^{\mu}=\left\{e^{-\eta_1(r)/2},\sqrt{1-\frac{r^2_0}{r^2}-\frac{r^2\Lambda}{3}},0,0\right\}.
\ee
Notice that the total effective stress-energy tensor is $T^{\text{eff}}_{\mu\nu} = T^{(\phi)}_{\mu\nu} - \frac{\Lambda}{\kappa^2} g_{\mu\nu}$. When contracted with a null vector $k^\mu$ (where $g_{\mu\nu}k^\mu k^\nu = 0$), the cosmological constant term vanishes entirely. Thus, the condition $T^{\text{eff}}_{\mu\nu} k^\mu k^\nu \geq 0$ strictly reduces to $T^{(\phi)}_{\mu\nu} k^\mu k^\nu \geq 0$. Nevertheless, the projection of the stress energy tensor along the null vectors reads
\be
T^{(\phi)}_{\mu\nu}k^\mu k^\nu=-\phi'^2\left(1-\frac{r^2_0}{r^2}-\frac{r^2\Lambda}{3}\right),
\ee
where we have utilized the fact that for the EB spacetime \cite{Ellis1973EtherFT, Bronnikov:1973fh}, the exotic scalar field only depends on the radial coordinate. Now, the quantity in parentheses remains positive throughout the interval $r_-<r\ll\sqrt{3/\Lambda}$ for the deformed EB WH constructed here. Taken together, these results indicate a violation of the null energy condition in the neighbourhood of the throat located at $r_-$, which is necessary to support the throat structure and thereby render the deformed EB WH solution traversable. Given the presence of an exotic matter field throughout the spacetime of the seed solution (the EB WH, see \eqref{eq: EB_action}), the violation of the null energy condition is an expected outcome for the deformed version. To make the traversable scenario more robust, we next compute the tidal force acting on the traveler.
\subsection{Tidal force on the traveler}
A generic feature in curved spacetime, as a manifestation of the curvature, is the deviation between two nearby geodesics. The effect is expected to be significant in the strong field regime, such as near the WH throat in the present context. The differential acceleration, commonly referred to as the tidal force, felt by a traveler while moving through the throat of the WH can be computed from the geodesic deviation equation, which, considering it in the traveler's locally inertial frame, in terms of the Riemann tensor, reads
\be
\Delta a^{\hat{i}}=-{R^{\hat{i}}}_{\hat{0}\hat{j}\hat{0}}\xi^{\hat{j}},
\ee
where $\Delta^{\hat{i}}$ stands for the differential acceleration, with ($\hat{i}$) in the superscript representing the spatial components in the orthonormal frame. Whereas $\xi$ denotes the separation between two points in the body of the traveler. Substituting the generic ansatz for the deformed WH \eqref{eq.2ndseed}, the radial and lateral components of the differential acceleration are given by \cite{morris1988wormholes}
\be
|\Delta a_{\rm rad}|=\left|\frac{e^{-\sigma(r)}}{2}\left[-\nu''(r)+\frac{\nu'(r)\sigma'(r)}{2}-\frac{\nu'^2(r)}{2}\right]\right||\xi|,
\ee
and 
\be
|\Delta a_{\rm lat}|=\gamma^2\left|\frac{e^{-\sigma(r)}}{2r}\left(v^2\sigma'(r)+\nu'(r)\right)\right||\xi|,
\ee
where $v$ represents the velocity of the traveler and $\gamma=\sqrt{1-v^2/c^2}$.  For the present deformed EB WH spacetime \eqref{eq:dsEBmetric}, the above expressions become
\be
\bea
|\Delta a_{\rm rad}|&=
\Bigg|
\frac{1}{2}\left(1-\frac{r^2_0}{r^2}-\frac{r^2\Lambda}{3}\right)
\Bigg\{
-\frac{2\Lambda}{3}\frac{r^2(r^2-3r_0^2)}{(r^2-r_0^2)^2}+\frac{2\Lambda^2 r^6}{9(r^2-r_0^2)^2}\Bigg\}
+\frac{\Lambda r^3\left(\frac{2r_0^2}{r^3}-\frac{2\Lambda r}{3}\right)}{3(r^2-r_0^2)}
\Bigg||\xi|
\eea
\ee
and 
\be\label{eq: lat_acc}
|\Delta a_{\rm lat}|=
\gamma^2
\left|\frac{\Lambda r^2\left(1-\frac{r^2_0}{r^2}-\frac{r^2\Lambda}{3}\right)}{3(r^2-r_0^2)}
+\frac{v^2}{r^2}
\left(
\frac{r_0^2}{r^2}-\frac{\Lambda r^2}{3}
\right)
\right||\xi|.
\ee
\begin{figure}[t!]
    \centering
\includegraphics[scale=0.55]{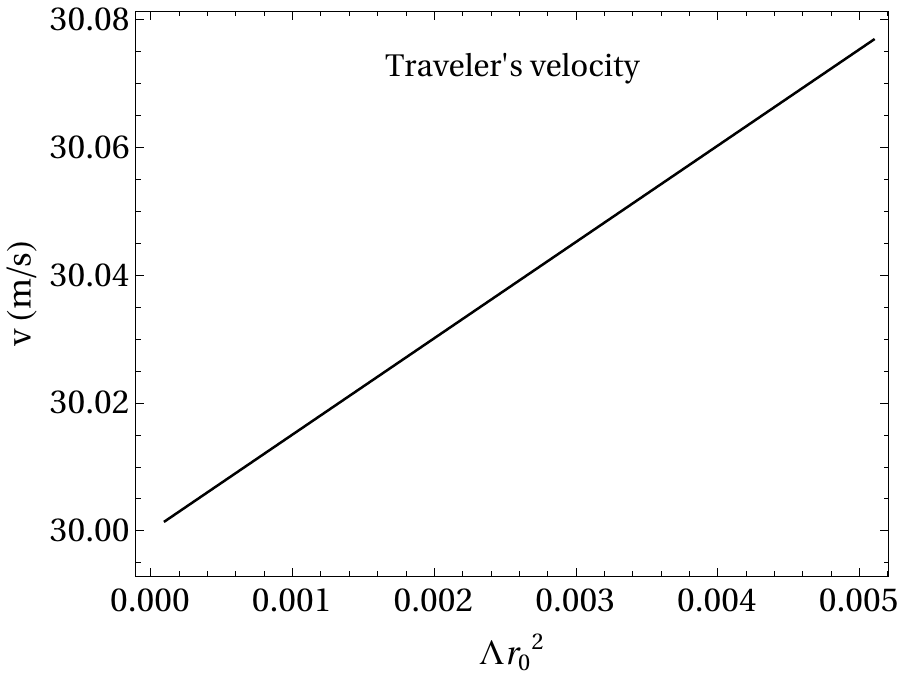}
\caption{The constraint on the velocity of the traveler near the throat of the deformed EB WH. Where we have set $r_0=10\,{\rm m}$.}\label{fig: velocity}
\end{figure}
As discussed in \Cref{sec: ds_soln}, for $\Lambda$ within $\Lambda<3/(4r^2_0)$, the deformed EB WH throat, $r_{-}$, lies outside $r_0$. Therefore, the above two tidal accelerations are finite everywhere, and the deformed EB WH should be traversable. Importantly, for the WH to be humanly traversable, the tidal accelerations must not exceed Earth’s gravitational acceleration, $g_\oplus$, i.e., $|\Delta a_{\rm rad}|<g_\oplus$ and $|\Delta a_{\rm lat}|<g_\oplus$. Therefore, these conditions can be further used to impose restrictions on the traveler's velocity and on the feasibility of human travel \cite{cataldo2017traversable, Zhang:2024kqp}. It is interesting to consider the limit $\Lambda\to 0$, asymptotically flat case first, where, the lateral tidal acceleration takes the following form, $v^2\gamma^2\le
r^2_0g_\oplus/|\xi|$. Where we have made use of the fact that at $\Lambda\to 0$, $r_{\rm th}\to r_0$, the restriction on the lateral acceleration is the most severe near the throat. For a realistic scenario, setting the size of a human traveler to be $|\xi|\sim 2\,{\rm m}$, yields $g_\oplus/|\xi|\sim 1/(10^{10}\, {\rm cm})^2$ in real unit. Choosing $r_0=10\, {\rm m}$, the constraint on the velocity turns out as $v\sim 30 {\rm m/s}$. On the other hand, considering non-zero $\lambda$, for non-relativistic motion, $v<<c$, $\gamma\sim 1$, the constraint on the velocity near the deformed EB WH throat, $r_-$, can be expressed as  
\be
v^2 \lesssim \frac{g_{\oplus}}{|\xi|}\left(
\frac{r^2}{\frac{r_0^2}{r^2}-\frac{\Lambda r^2}{3}}\right)\Bigg|_{r=r_-}. 
\ee 
In Fig.~\ref{fig: velocity}, we have plotted this velocity with respect to $\Lambda$. With a closer look to \eqref{eq: lat_acc}, it can be realized that, in the presence of the positive cosmological constant, the lateral component of the tidal acceleration is less severe near the throat as compared to the asymptotically flat case, which leads to a higher velocity for the traveler. Physically, it represents the maximum allowed velocity with the acceleration $g_{\oplus}$. However, it should be noted that the above expression, and the corresponding numerical values of the velocity, do not provide a uniform constraint (independent of $r$ in the region: $r_-<r<\sqrt{3/\Lambda}$) as found in the case of asymptotically flat case. For this reason, we find it difficult to compute the time taken by the traveller during the trip, which we leave for future investigation. Nevertheless, the analysis reveals that tidal forces remain finite throughout the region of interest, while the derived upper bound on the velocity, normalized to Earth’s gravitational acceleration, ensures compatibility with non-disruptive motion of a traveler, thereby supporting the physical consistency of traversability within the spacetime. We next consider the geodesic motion in the present deformed WH background, to check whether it is possible to extend it through the throat.
\subsection{Geodesics of massless particles in the deformed Ellis-Bronnikov wormhole spacetime}
The Lagrangian density describing geodesic motion of a particle confined to the equatorial plane in the static, spherically symmetric spacetime \eqref{eq.2ndseed} is given by
\be
2\mathcal{L}=-e^{\nu(r)}\dot{t}^2+e^{\sigma(r)}\dot{r}^2+r^2\dot{\phi}^2,
\ee
where the overdot indicates differentiation with respect to an affine parameter, which for massive particles may be identified with the proper time. The components of the generalized momenta associated with the particle can be obtained from the above Lagrangian and can be expressed as :
\be
\bea
p_t&=-e^{\nu(r)}\dot{t}=-E,\\
p_r&=e^{\sigma(r)}\dot{r}\\
p_\phi&=r^2\dot{\phi}=L,
\eea
\ee
where $E$ and $L$ denote the conserved energy and angular momentum, respectively, as the spacetime allows for $\pr_t$ and $\pr_\phi$ killing vectors. Then the geodesics are described by $p_\mu\dot{x}^\mu=-m$. For example, the motion of the particles with $m=0$ are governed by null geodesics, whereas, for $m\neq 0$, they are time-like geodesics. With the help of the above set of generalized momenta, the equation corresponding to the radial motion reads
\be
\dot{r}^2+V_{\rm eff}(r)=0,
\ee
with 
\be
V_{\rm eff}(r)=\left\{\begin{matrix}
    e^{-\sigma(r)}\left(-\frac{E^2}{e^{\nu(r)}}+\frac{L^2}{r^2}\right),~~~~~m=0,\\
     \frac{e^{-\sigma(r)}}{m^2}\left(m^2-\frac{E^2}{e^{\nu(r)}}-\frac{L^2}{r^2}\right),~~~~~m\neq 0.
\end{matrix}
\right.
\ee
For simplification, we consider $L=0$ and $m=0$, and study the motion along null radial geodesics. Then the radial equation can be expressed as 
\be
\left(\frac{dr}{d\lambda}\right)^2-e^{-\sigma(r)-\nu(r)}E^2=0,
\ee
where $\lambda$ stands for the affine parameter associated with the geodesics. In what follows, we will analyze the evolution of the affine parameter as the geodesic approaches the vicinity of the throat. For this purpose, the above equation should be integrated as
\be
\tilde{\lambda}-\tilde{\lambda}_0=\pm \int^r_{r_0}e^{\sigma(r')+\nu(r')}\,dr',
\ee
where the original affine parameter has been rescaled as $\tilde{\lambda}=\lambda/E$, with $\tilde{\lambda}_0$ denoting the value of the rescaled affine parameter at the throat. Inserting the analytical solution of redshift and the shape function for the present deformed EB WH from \eqref{eq:dsEBmetric}, we get
\be\label{eq: affine_int}
\tilde{\lambda}-\tilde{\lambda}_0=\pm \int^r_{r_-}\frac{({r'}^2-r^2_0)^{-\Lambda r^2_0/6}e^{-\Lambda r'^2/6}}{\sqrt{1-\frac{r^2_0}{r'^2}-\frac{r'^2\Lambda}{3}}}\,dr'.
\ee
Let us analyze the behaviour of the integrand near the WH throat, $r_-$, where we have
\be
\lim_{r'\to r_-} \frac{({r'}^2-r^2_0)^{-\Lambda r^2_0/6}e^{-\Lambda r'^2/6}}{\sqrt{1-\frac{r^2_0}{r'^2}-\frac{r'^2\Lambda}{3}}}\sim \frac{({r}^2_--r^2_0)^{-\Lambda r^2_0/6}}{\sqrt{r'-r_-}}.
\ee
The above limit indicates that, for the present deformed EB WH, having $r_->r_0$, the integral \eqref{eq: affine_int} is finite in the vicinity of the throat, implying that a massless particle can reach the throat within a finite value of the affine parameter. Moreover, at $r'=r_-$, the expression for the affine parameter \eqref{eq: affine_int} exhibits an integrable divergence, which indicates that the affine parameter can be smoothly extended across the throat.
\begin{figure}[t!]
    \centering
\includegraphics[scale=0.50]{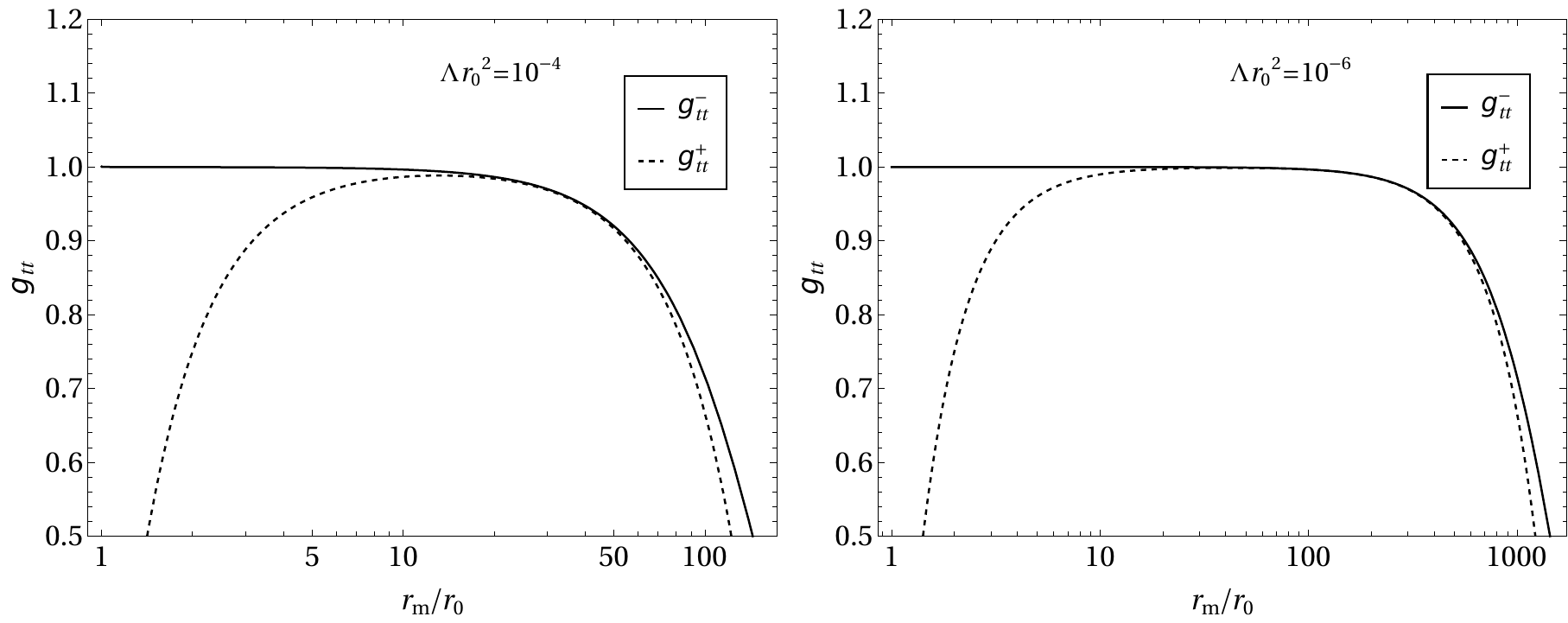}
\caption{The $g_{tt}$ component of both the interior ($g^-_{tt}$ of deformed EB WH) and the exterior ($g^+_{tt}$ of Schwarzschild de-Sitter) has been plotted with the radial distance. Variations have been shown for two different values of the cosmological constant with $\Lambda r^2_0\ll 1$. The overlapping points can be identified as the matching points suitable to join the two line elements.}\label{fig: matching_point}
\end{figure}
\section{Extension to a Schwarzschild-de Sitter exterior spacetime}\label{sec: exterior}
The discussion so far has been restricted to the deformed EB WH, whose perturbative description is valid only within a finite radial domain, $r\ll\sqrt{3/\Lambda}$. A physically viable spacetime, however, must possess a well-defined exterior region and asymptotic structure, which are also essential for astrophysical applications. It is therefore imperative to construct a global extension by matching the locally deformed geometry to an appropriate exterior solution at a finite radius. Having considered a static, spherically symmetric WH geometry in the presence of the positive cosmological constant, it is natural to take the de Sitter-Schwarzschild spacetime as the exterior solution. The matching radius, $r_{\rm m}$, is chosen to lie outside the de Sitter-Schwarzschild event horizon (so that it does not represent a black hole).  In general, such a matching is performed using the Israel junction conditions \cite{Israel1966}. For the present spherically symmetric spacetime with fixed source and restricted radial domain, we adopt a simplified approach and leave a more rigorous treatment for future work. Specifically, following Ref. \cite{lemos2003morris}, we directly match the interior and exterior metric functions at the junction surface. The de Sitter-Schwarzschild line element is given by 
\be
ds^2_+=-\left(1-\frac{2M}{r}-\frac{\Lambda r^2}{3}\right)dt^2+\frac{dr^2}{1-\frac{2M}{r}-\frac{\Lambda r^2}{3}}+r^2d\theta^2+r^2\sin^2\theta d\varphi^2,
\ee
where $M$ denotes the Schwarzschild mass parameter, and the $+$ sign in the subscript is just for labelling it as the exterior spacetime. The angular part of the above exterior spacetime is already continuous with the interior \eqref{eq:dsEBmetric}. Therefore, we are left with matching the redshift and the shape function at the boundary radius $r_{\rm m}$, for $r_-<r_{\rm m}\ll\sqrt{3/\Lambda}$. Matching the shape functions yields
\be
1-\frac{2M}{r_{\rm m}}-\frac{\Lambda r^2_{\rm m}}{3}=1-\frac{r^2_0}{r^2_{\rm m}}-\frac{r^2_{\rm m}\Lambda}{3}.
\ee
From this relation, the mass parameter is obtained as $M=r^2_0/2r_{\rm m}$. On the other hand, matching the redshift functions gives 
\be
e^{-\frac{\Lambda}{3}\,r^2_{\rm m}-\frac{\Lambda r_0^2}{3}\,\ln\!\left|\frac{r^2_{\rm m}}{r_0^2}-1\right|}=1-\frac{2M}{r_{\rm m}}-\frac{\Lambda r^2_{\rm m}}{3}.
\ee
Since this condition is transcendental in $r_{\rm m}$, it is not straightforward to determine analytically whether a physically acceptable matching radius exists. We therefore analyze by numerically plotting the two sides of the above to identify the intersection points. Also, we require that $r_{\rm m}>2M$, which, using the expression for the mass parameter obtained above, implies $r_{\rm m}>r_0$. Since the throat of the deformed EB WH satisfies $r_->r_0$, this condition is automatically satisfied provided that the matching radius lies in the physically admissible domain $r_-<r_{\rm m}\ll\sqrt{3/\Lambda}$. In this regime, Fig.~\ref{fig: matching_point} demonstrates that a significant overlap region exists for determining the matching radius, allowing a smooth connection between the interior and exterior spacetimes and thereby providing the complete global extension of the deformed EB WH. 
\section{Conclusion and outlook}\label{sec: conclusion}
In this work, we have studied the deformations of asymptotically flat WH solutions in the presence of a positive cosmological constant in a perturbative manner. Starting from the Ellis–Bronnikov seed geometry, we have shown that the preservation of spherical symmetry, together with the uniform nature of the cosmological constant term in the Einstein field equations, naturally facilitates the application of the perturbative method. By treating the cosmological constant as an effective source term and shifting it to the matter sector, the resulting deformation can be consistently determined without introducing additional assumptions, apart from the smallness of the cosmological constant, $\Lambda r^2_0\ll 1$. However, the perturbative deformation remains valid within a restricted radial domain $r\ll\sqrt{3/\Lambda}$. Therefore, to extend the locally deformed geometry beyond this regime, we consider a matching with a Schwarzschild–de Sitter exterior spacetime at a suitable matching radius, which is determined numerically from the matching conditions.

Despite the deformation induced by the cosmological constant, the flare-out condition remains satisfied in the vicinity of the throat. Furthermore, the finiteness of the tidal forces and the regular evolution of the affine parameter along geodesics confirm the absence of singular behaviour. These results indicate that the essential traversable WH structure of the original Ellis–Bronnikov WH geometry is preserved at the perturbative level. In the existing literature on the construction of thin-shell WHs via the cut-and-paste technique \cite{lemos2003morris}, the effect of the cosmological constant enters primarily through the exterior geometry and the matching procedure at the junction surface, while its role in the WH interior is reflected mainly in reducing the amount of NEC-violating matter required to support the WH. In contrast, in our approach, the cosmological constant is incorporated directly into the WH interior geometry, allowing us to investigate its effect on the throat structure and the resulting deformation of the seed WH solution.

Finally, the methodology developed here for computing deformations in WH geometries in the presence of a positive cosmological constant can be straightforwardly extended to other asymptotically flat, static, and spherically symmetric WH solutions. For example, generalized Ellis–Bronnikov configurations, which belong to the Morris–Thorne class \cite{Kar:1995jz}, can be directly analyzed within the present framework. However, for WH geometries with a non-vanishing redshift function, such as the Damour–Solodukhin WH \cite{Damour:2007ap}, one may encounter additional difficulties in integrating the decoupled Einstein equations. Nevertheless, the present formalism provides a simplified and systematic framework to explore such extensions. Since current observational probes, such as gravitational wave measurements and black hole shadow image observation by the Event Horizon Telescope (EHT), are primarily sensitive to the spacetime geometry up to the photon sphere, the possibility remains open that exotic compact objects could coexist with the vacuum solutions \cite{LIGOScientific:2021sio, Vagnozzi:2022moj}. In this context, a natural first step is to investigate the stability of the present WH solution under perturbations by various fundamental fields, including scalar, electromagnetic, and gravitational perturbations. Such an analysis would further enable the computation of the corresponding quasinormal mode spectrum and ringdown waveform, allowing for a direct comparison with the templates that currently exist for vacuum black hole spacetimes \cite{Azad:2022qqn, DuttaRoy:2019hij}. Moreover, the gravitational lensing properties and shadow images associated with the present WH geometry can be explored in detail. It would also be interesting to investigate how the presence of a cosmological constant modifies these observational signatures relative to those of the standard Ellis–Bronnikov WH.

\textbf{Acknowledgements :} We would like to thank Mir Afrasiar for various helpful discussions. We also acknowledge Wonwoo Lee for providing insightful comments on the manuscript. This work
was partially supported by the National Natural Science Foundation of China (NSFC) (Grant Nos. 12275166
and 12311540141). 
\bibliographystyle{JHEP}
\bibliography{ref} 
\end{document}